\documentclass[aps,twocolumn,superscriptaddress,showpacs]{revtex4}
\input epsf

\begin{document}

\title{Computer simulations of liquid silica: equation of state and
liquid-liquid phase transition}

\author{Ivan Saika-Voivod}
\affiliation{Department of Applied Mathematics,
University of Western Ontario, London, Ontario N6A~5B7, Canada}

\author{Francesco Sciortino}
\affiliation{Dipartimento di Fisica and Istituto Nazionale per la
Fisica della Materia, Universita' di Roma La Sapienza, Piazzale Aldo
Moro~2, I-00185, Roma, Italy}

\author{Peter H. Poole}
\affiliation{Department of Applied Mathematics,
University of Western Ontario, London, Ontario N6A~5B7, Canada}

\date{\today}

\begin{abstract}
We conduct extensive molecular dynamics computer simulations of two
models for liquid silica [the model of Woodcock, Angell and Cheeseman,
J. Phys. Chem. {\bf 65}, 1565 (1976); and that of van~Beest, Kramer
and van~Santen, Phys. Rev. Lett. {\bf 64}, 1955, (1990)] to determine
their thermodynamic properties at low temperature $T$ across a wide
density range.  We find for both models a wide range of states in
which isochores of the potential energy $U$ are a linear function of
$T^{3/5}$, as recently proposed for simple liquids [Rosenfeld and
P. Tarazona, Mol. Phys. {\bf 95}, 141 (1998)].  We exploit this
behavior to fit an accurate equation of state to our thermodynamic
data.  Extrapolation of this equation of state to low $T$ predicts the
occurrence of a liquid-liquid phase transition for both models.  We
conduct simulations in the region of the predicted phase transition,
and confirm its existence by direct observation of phase separating
droplets of atoms with distinct local density and coordination
environments.
\end{abstract}

\pacs{65.50.+m 64.30.+t, 64.60.My, 64.70.Ja}
\maketitle

\section{Introduction}

First order liquid-liquid phase separation, in which two liquids of
distinct chemical composition coexist, are common in multicomponent
systems.  However, there has in recent years been a growing interest
in first order liquid-liquid phase transitions that occur without a
change of composition, but rather with a change in density $\rho$ as
temperature $T$ or pressure $P$ is varied.  Experimental evidence for
the occurrence of such transitions has been found for a wide range of
systems, including Si~\cite{T84,DSTPJ85}, I, Se,
S~\cite{BVPU91,BPVU92}, Al$_2$O$_3$-Y$_2$O$_3$ melts~\cite{AM94},
C~\cite{T97}, H$_2$0~\cite{MS98,M00}, and P~\cite{K00}.  Liquid-liquid
transitions have also been observed in molecular dynamics (MD)
computer simulations of Si~\cite{LL89,ABG96},
H$_2$O~\cite{PSES92,HZPSS97} and C~\cite{GR99}.  Theoretical studies
have long predicted liquid-liquid transitions for a variety of model
fluids; for examples, see
Refs.~\cite{HS70,MPS85,PB92,SSS93,PSGSA94,RPD96,TB98,J99,FMSBS00,J00}.

In the case of water, the proposed liquid-liquid phase transition
occurs in the supercooled liquid, i.e. for $T$ less than that of the
melting line~\cite{MS98}.  Closely associated with the possibility of
a liquid-liquid transition in supercooled water is the phenomenon of
polyamorphism in the amorphous solid occurring below the glass
transition temperature, $T_g$.  Polyamorphism refers to the occurrence
of distinct amorphous solid forms of a material~\cite{W92,PGAM97}.  In
the most prominent cases of polyamorphism, such as
water~\cite{MCW85,MTA91,M94}, an abrupt first-order-like transition
occurs from a low-density form to a distinct high-density form as the
amorphous material is compressed at low $T$.  For water it was
proposed that the observed polyamorphism of the amorphous solid is a
sub-$T_g$ manifestation of the thermodynamic instability associated
with the liquid-liquid phase transition~\cite{PSES92}.

Computer simulation studies of the ST2 water model~\cite{SR74} support
this view of the relationship between liquid-liquid phase transitions
and polyamorphism.  The qualitative features of water polyamorphism
are clearly displayed in simulations of ST2 water~\cite{PESS93}.
Correspondingly, the critical temperature $T_c$ marking the onset of
liquid-liquid phase separation has been determined~\cite{HZPSS97}, as
well as a characteristic pattern of thermodynamic ``precursors'' for
$T>T_c$ consistent with the instability at $T=T_c$~\cite{SPES97}.
These precursors include the occurrence of a density maximum and a
compressibility maximum. Simulation studies of the thermodynamic
properties of two other water-like models, TIP4P~\cite{JCMIK83} and
SPC/E~\cite{B84}, find the same pattern of thermodynamic precursors
observed for ST2 for $T>T_c$~\cite{HPSS97,SPES97}, and also display
polyamorphism in simulations of the amorphous solid.  However, in
these systems, simulations of the liquid at lower $T$, to test for the
onset of a liquid-liquid phase transition, have not yet been attempted
due to prohibitively long equilibration times.

Substances that are structurally similar to water, such as Si, Ge,
GeO$_2$, and SiO$_2$ (silica), have the potential to exhibit similar
behavior~\cite{P98}.  Particular attention has been paid to silica
because of its technological and geological importance.  Polyamorphism
is indeed observed in compression experiments on amorphous
silica~\cite{SKS82,G84,HMBM86}, and is also qualitatively reproduced
in computer simulations~\cite{SSGCS93,TKL92,JKV93}.  Though not as
dramatic as is found for amorphous solid water, the polyamorphism of
silica may also be due to a trend toward liquid-liquid phase
separation~\cite{PHA97,L00}.  Indeed, liquid state simulations of the
silica model of Woodcock, Angell and Cheeseman (denoted here ``WAC
silica'')~\cite{WAC76} have shown that the same pattern of
thermodynamic precursors of the liquid-liquid phase transition found
in water simulations also occurs in this system~\cite{PHA97}. However,
as in TIP4P and SPC/E water, the low $T$ simulations required to test
for an explicit liquid-liquid phase transition in WAC silica have not
been attempted to date.

At the same time, recent advances concerning the properties of liquids
at and below the melting line are expanding our ability to study the
states where liquid-liquid phase transitions may occur.  Of particular
importance is the recent prediction~\cite{RT98} that an isochore of
the potential energy $U$ should be a linear function of $T^{3/5}$ for
a simple, cold, dense liquid.  Since this relation is proposed to be
valid in the limit of low $T$, it is a useful relation for studying
the properties of a deeply supercooled liquid.  Notably, several
recent works showed that this prediction is obeyed at low $T$ for
binary Lennard-Jones liquids~\cite{prlentro,barbara,srinew}.  This
observation provided a physical basis for extrapolating $U$ (and
thermodynamic properties derived from it) to $T$ near $T_g$, and so
made possible determination of the Kauzmann temperature for this
system.

In this paper, we examine in detail the behavior of two simulation
models of silica: (i) WAC silica; and (ii) the widely used potential
of van~Beest, Kramer and van~Santen~\cite{BKS90}, denoted here as
``BKS silica.'' Our goal is to determine if either model displays a
liquid-liquid phase transition.  We find that our computer simulation
data for the thermodynamic properties of these silica models obey the
prediction of Ref.~\cite{RT98} over a wide range of $T$ and $V$.  We
exploit this result to construct equations of state for BKS and WAC
silica, and find that liquid-liquid phase transitions are predicted
for both models at low $T$.  We then conduct simulations near the
predicted phase transition, and confirm the occurrence of
liquid-liquid phase transitions for both BKS and WAC silica by direct
observation.

\section{Molecular Dynamics Simulations}

The calculations presented here for WAC silica are based on the data
set generated for Ref.~\cite{PHA97}.  These simulations consisted of
$N=450$ atoms (300 O, 150 Si atoms) and were conducted in the constant
$(N,V,E)$ ensemble.  ($E$ is the total internal energy, $V$ is the
volume.)  The effects of electrostatic interactions were incorporated
using the Ewald summation technique~\cite{AT89}.

For the BKS model, we conduct new simulations of a system of $N=1332$
atoms.  As for WAC silica, Ewald summations are used to include
electrostatic interactions.  The Ewald parameter ($\alpha$ in the
notation of Ref.~\cite{AT89}) is fixed to 2.5~nm$^{-1}$ for all state
points simulated.  For each state point, the system is equilibrated to
near the desired $T$ using periodic velocity rescaling.  All averages
are reported for constant $(N,V,E)$ simulations that follow the
equilibration stage.  In all cases, averages are evaluated over a time
that is at least ten times longer than the average time required for
an Si atom to diffuse $0.2$~nm.

The BKS potential has the unphysical feature that the interaction
energy of a Si and O atom pair diverges to $-\infty$ as their
separation goes to $0$.  Though not a problem at ambient $T$ and $P$,
this feature will occasionally manifest itself at high $T$ and $P$.
We have added a short range term to the BKS potential that prevents
this from occurring, but which does not alter the form of the BKS
potential at larger separations~\cite{mod}.

The $(V,T)$ coordinates of the state points simulated for this work
are shown in Fig.~\ref{maps}.  Table~\ref{vlist} gives $V$ for each of
the isochores studied.

\section{Temperature dependence of potential energy isochores}

Ref.~\cite{RT98} predicts that the isochoric $T$ dependence of the
potential energy $U$ of a simple, dense, cold liquid is given by,
\begin{equation}
U=a+b\,T^{3/5},
\label{tara1}
\end{equation}
where $a$ and $b$ are constants for a given $V$.  

\begin{figure}
\hbox to\hsize{\epsfxsize=1.0\hsize\hfil\epsfbox{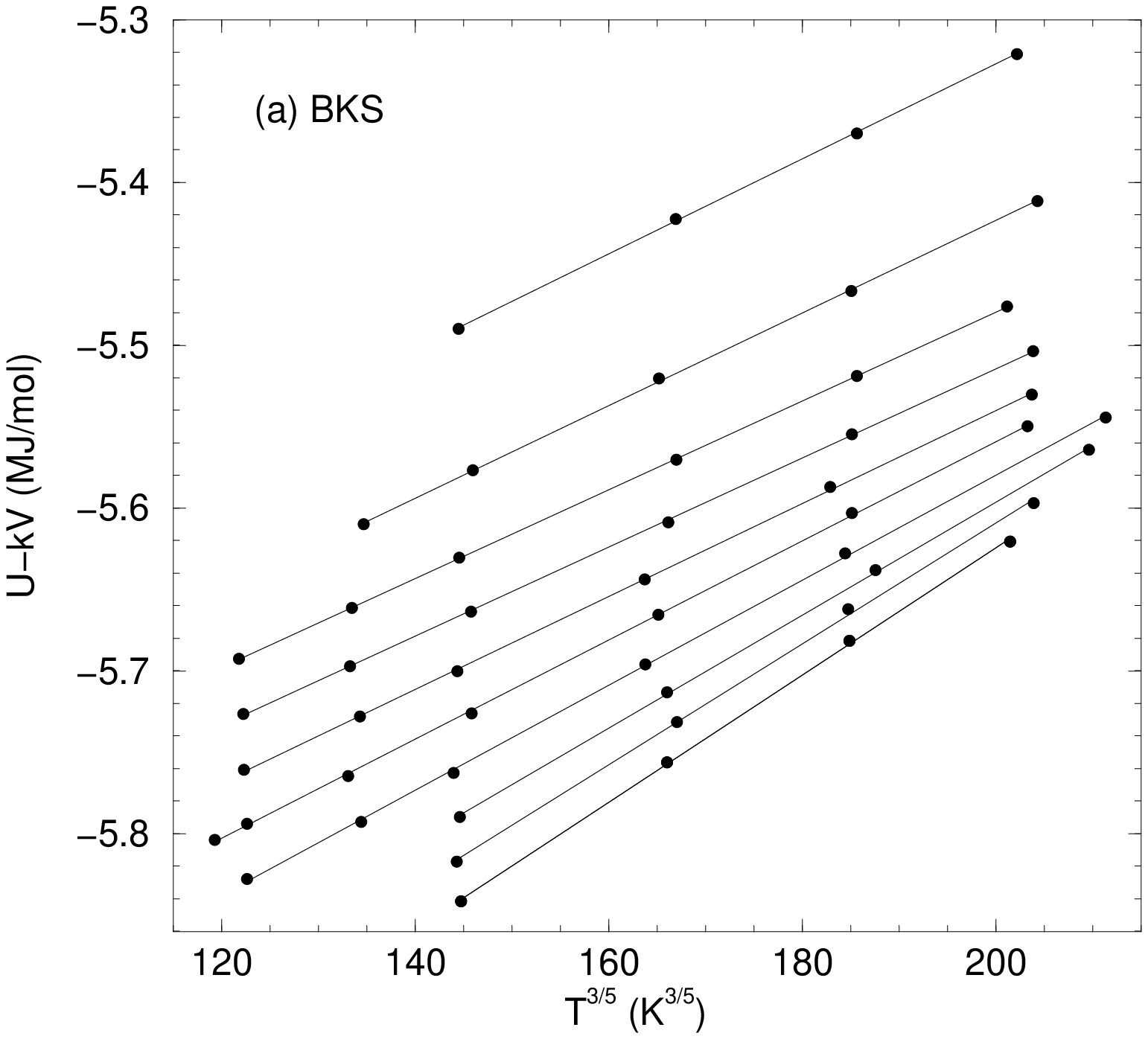}\hfil}
\hbox to\hsize{\epsfxsize=1.0\hsize\hfil\epsfbox{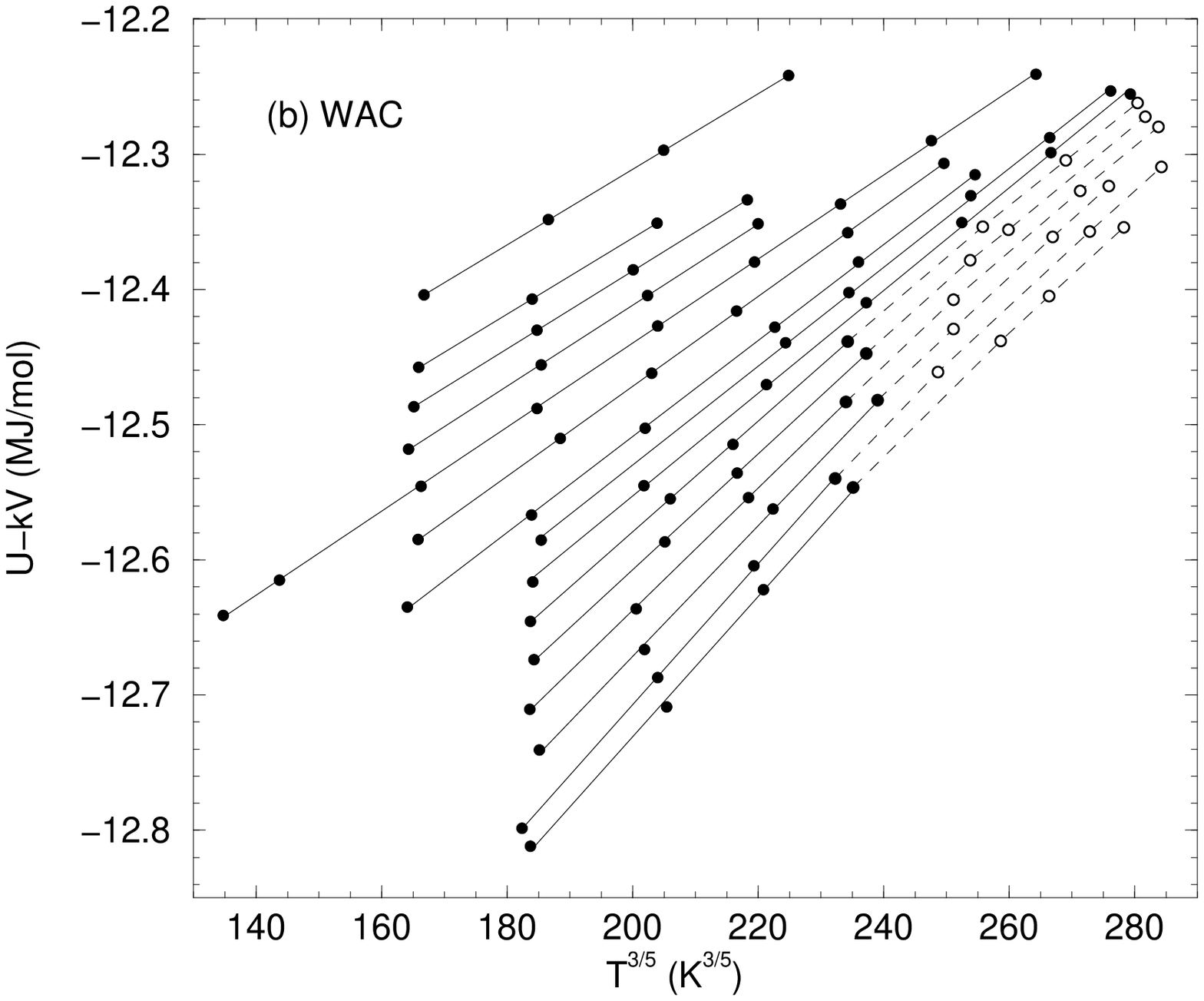}\hfil}
\caption{Isochores of $U$ versus $T^{3/5}$ for (a) BKS and (b) WAC
silica.  Symbols are $U$ values obtained from MD simulation, while
solid lines show linear fits to each isochore.  To better view each
isochore and their fitted lines, we plot $U-kV$, with
$k=1$~(g\,MJ)/(cm$^3$\,mol) and $V$ measured in cm$^3$/g, so that each
isochore is subject to a $V$-dependent shift to separate it from the
others.  In (a) isochores B1 through B10 are shown from bottom to top;
in (b) isochores W1 through W15 are shown from bottom to top.  Also in
(b) are shown high $T$ points (open circles connected by dashed lines)
on the largest $V$ isochores which deviate from linear behavior and so
are excluded from the fits.  The statistical error for the $U$ values
shown in (a) and (b) does not exceed $\pm 0.004$~MJ/mol.}
\label{UT}
\end{figure}

Here we test if Eq.~\ref{tara1} is obeyed by WAC and BKS silica.  We
plot isochores of $U$ against $T^{3/5}$ and fit a straight line to the
data (Fig.~\ref{UT}).  At all $V$ studied, we find that
Eq.~\ref{tara1} fits the data well.  Consistent with the prediction of
Ref.~\cite{RT98}, the best fits occur for the smallest $V$, and the
quality of the fits decreases somewhat as $V$ increases. For BKS
silica, we find that Eq.~\ref{tara1} fits to all isochores within
numerical uncertainty.  For WAC silica, most of the data available to
us fits Eq.~\ref{tara1} within numerical uncertainty.  However, for
the largest $V$, systematic deviations from Eq.~\ref{tara1} occur if
all WAC data up to the highest $T$ are included.  Yet even for these
large $V$ isochores, the WAC data are consistent with an approach to
the behavior of Eq.~\ref{tara1} at low $T$.  By excluding several of
the highest $T$ data points, shown as open circles in Fig.~\ref{maps},
we recover a fit within numerical uncertainty even for the largest $V$
isochores for WAC silica.

\begin{figure}
\hbox to\hsize{\epsfxsize=1.0\hsize\hfil\epsfbox{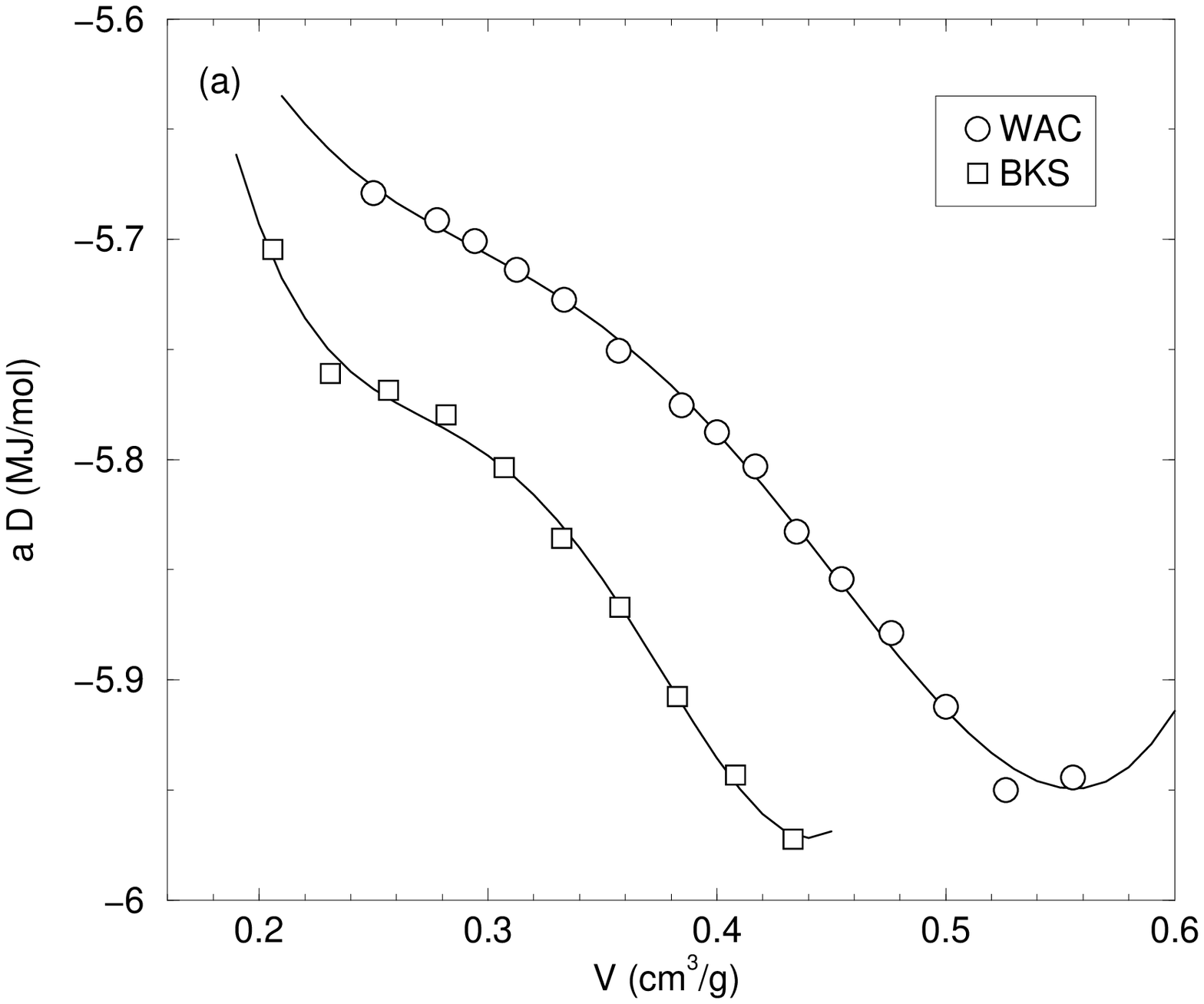}\hfil}
\hbox to\hsize{\epsfxsize=1.0\hsize\hfil\epsfbox{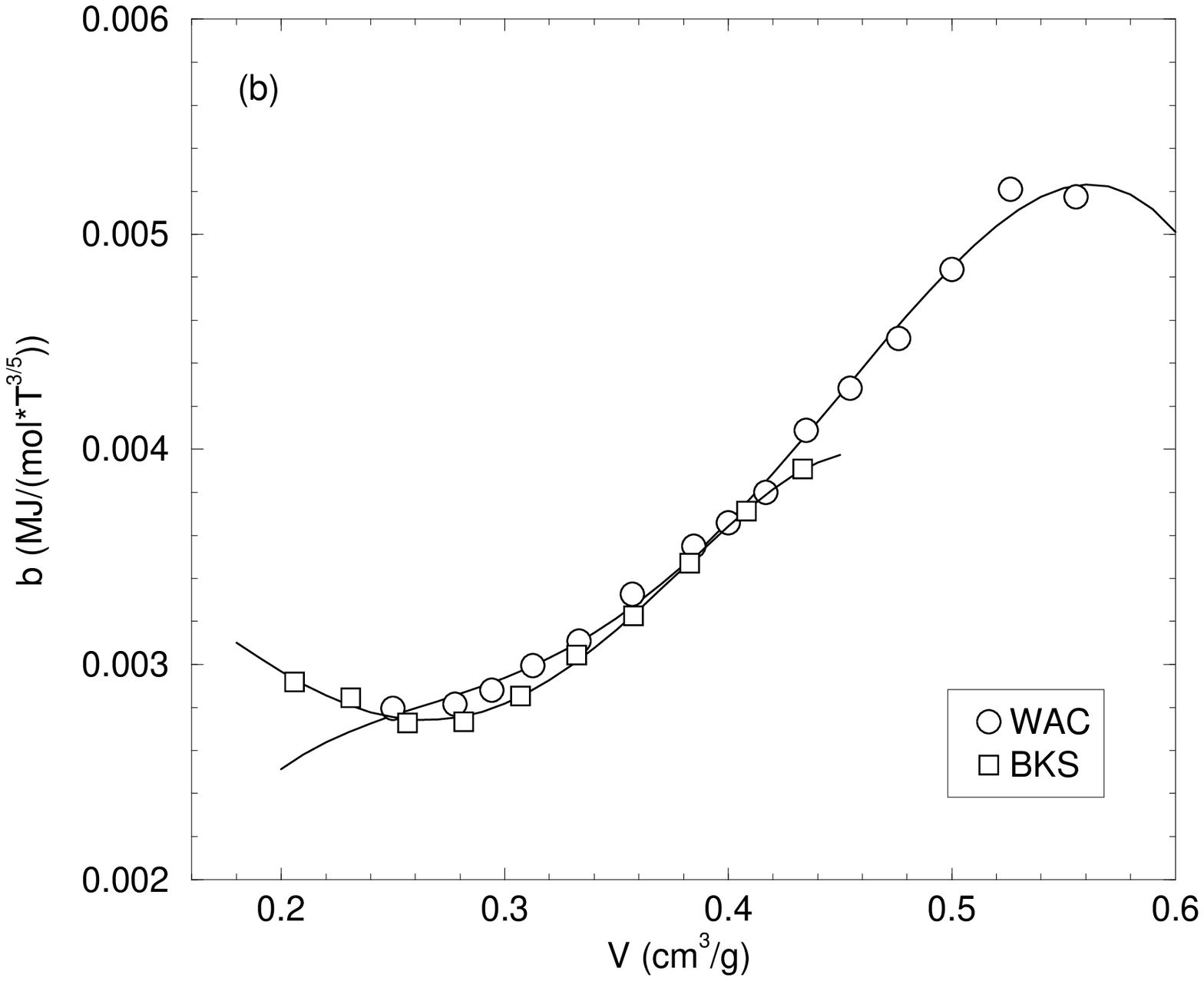}\hfil}
\caption{$V$ dependence of (a) $a$ and (b) $b$ for BKS and WAC silica,
found from the linear fits to the isochores shown in
Fig.~\protect{\ref{UT}}.  Solid curves are fits to the data of a
fourth order polynomial in $V$.  In (a) $D$ is a scale factor to
permit both curves to be compared in a single plot; for BKS, $D=1$ and
for WAC, $D=0.45$.}
\label{ab}
\end{figure}

The $V$ dependence of the fit parameters $a$ and $b$ so obtained for
both models is shown in Fig.~\ref{ab}.  The success of the fits to
Eq.~\ref{tara1} over a wide range of $T$ and $V$, and the smooth
variation of $a$ and $b$ with $V$ show that the predictions of
Ref.~\cite{RT98} appear to be valid for silica, a ``complex'' liquid
with anisotropic molecular interactions, at least in the limits of low
$T$ and low $V$.

In addition, it is important to note the $V$ dependence of $a$. $a$
provides an estimate of $U$ in the limit $T\to 0$, which classically
is coincident with the limit as $T\to 0$ of $A$, the Helmholtz free
energy.  For both WAC and BKS silica we find a range of $V$ in which
the curvature of the $a$ versus $V$ curve is negative.  In this range,
the condition for thermodynamic stability for a single phase,
$(\partial^2 A/\partial^2 V)_T>0$, is not satisfied~\cite{Callen},
suggesting that both WAC and BKS silica would undergo a liquid-liquid
phase separation at low $T$, if not preempted by crystallization or
vitrification.

\section{Model Equation of State}
\label{model}

Having identified a region of validity for Eq.~\ref{tara1}, we can use
this relation to construct a representation of the thermodynamic
properties of WAC and BKS silica in terms of a continuous function of
$T$ and $V$.  The model equations of state so generated will allow us
to clarify the liquid-liquid phase separation suggested by the $V$
dependence of $a$ in Fig.~\ref{ab}.

We first fit polynomials to the $V$ dependence of $a(V) \simeq
\sum_{n=0}^{4} \alpha_n V^n$ and $b(V) \simeq\sum_{n=0}^{4} \beta_n
V^n$, to obtain a functional representation of $U$:
\begin{equation}
U(V,T)=a(V)+b(V)\,T^{3/5}.
\label{tara}
\end{equation}
The coefficients $\alpha_n$ and $\beta_n$ are given in
Table~\ref{coeffs}.  The internal energy is $E=U+U_k$; for the
classical ionic models of silica considered here, the kinetic energy
is $U_k=\frac{9}{2}RT$, where $R$ is the gas constant.  Hence, our
model for $E$ is,
\begin{equation}
E(V,T)=a(V)+b(V)\,T^{3/5} + \frac{9}{2}RT.
\label{E}
\end{equation}

Next, we seek a functional representation of the entropy $S(V,T)$, so
that a model for $A(V,T)$ can be obtained from $A=E-TS$.  $S$ at
arbitrary $V$ and $T$, relative to the entropy $S(V_0,T_0)$ of a
reference state, can be evaluated by thermodynamic
integration~\cite{Callen}.  We carry this out in two steps, first
along an isotherm, and then along an isochore.

\begin{figure}
\hbox to\hsize{\epsfxsize=1.0\hsize\hfil\epsfbox{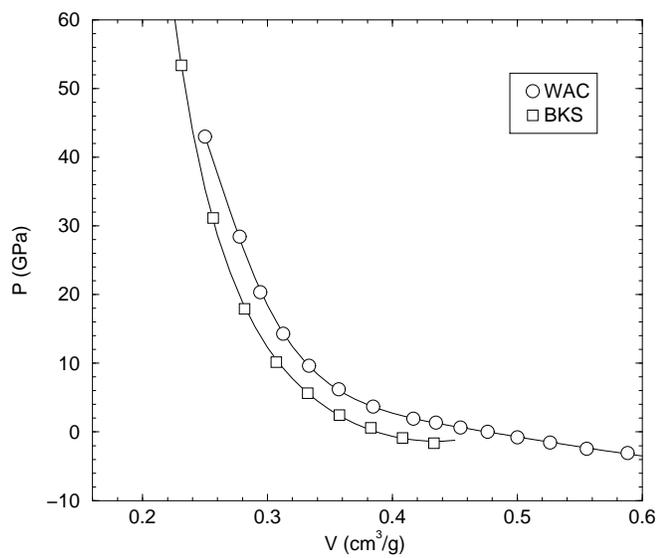}\hfil}
\caption{Isotherms of $P$ versus $V$ for BKS ($T_0=5000$~K) and WAC
($T_0=7000$~K) silica.  Solid curves are fits to the data of a fifth
order polynomial in $\rho$.}
\label{PV}
\end{figure}

We compute the change $\Delta S_T = S(V,T_0)-S(V_0,T_0)$ along an
isotherm $T=T_0$ using $\Delta A=\Delta E-T_0\Delta S_T$, where
$\Delta A = A(V,T_0)-A(V_0,T_0)$ and $\Delta E = E(V,T_0)-E(V_0,T_0)$.
$\Delta E$ is evaluated from Eq.~\ref{E}.  $\Delta A$ is a
difference due to a volume change from $V_0$ to $V$ at $T=T_0$, and is
found from an isothermal integration of $dA=-PdV$; that is, 
\begin{equation}
\Delta A=-\int_{V_0}^V P(V',T_0) \,dV'.
\label{A}
\end{equation}
To obtain a functional representation for $\Delta A$ therefore
requires one for $P(V)$ at $T=T_0$. We obtain the required data from
our MD simulations and fit to them a polynomial $P\simeq
\sum_{n=0}^{5} \gamma_n \rho^n$, where the density $\rho=1/V$
(Fig.~\ref{PV}).  The coefficients $\gamma_n$ are given in
Table~\ref{coeffs}, along with the choices of $T_0$ for WAC and BKS
silica.  Using this polynomial model for $P(V,T_0)$, the integration
in Eq.~\ref{A} yields an expression for $\Delta A$ which combined with
that for $\Delta E$, gives a model function for $\Delta
S_T$~\cite{maple}.  In terms of $E$ and $P$, the expression for
$\Delta S_T$ is,
\begin{equation}
      \Delta S_T=\frac{1}{T_0}\biggl[ E(V,T_0)-E(V_0,T_0)
      +\int_{V_0}^V P(V',T_0)\,dV' \biggr].
\end{equation}

The change $\Delta S_V = S(V,T)-S(V,T_0)$ is the entropy difference at
fixed $V$ due to a temperature change from $T_0$ to $T$.  This we
find from an isochoric integration of,
\begin{equation}
dS=\frac{dE}{T}=\frac{1}{T}
\biggl(\frac{\partial E}{\partial T}\biggr)_V \,dT.
\end{equation}
That is, 
\begin{equation}
      \Delta S_V=\int_{T_0}^{T} \frac{1}{T'} \biggl(\frac{\partial
E}{\partial T'}\biggr)_{V} \, dT'.
\end{equation}
This evaluation is carried out using our representation of $E(V,T)$ in
Eq.~\ref{tara}.

Combining the contributions of both isothermal and isochoric changes,
$S$ at an arbitrary state point is given by,
\begin{equation}
S(V,T)=S(V_0,T_0)+\Delta S_T+\Delta S_V.
\end{equation}

Using the model functions for $E$ and $S$, we thus obtain a function
modeling $A(V,T)$.  The equation of state $P(V,T)$ is found from,
\begin{equation}
P(V,T)=-\biggr(\frac{\partial A}{\partial V}\biggl)_T.
\label{PVT}
\end{equation}
Note that the resulting expression for $P(V,T)$ does not contain the
unknown reference entropy $S(V_0,T_0)$ since this constant disappears
after the differentiation in Eq.~\ref{PVT}.

\begin{figure}
\hbox to\hsize{\epsfxsize=1.0\hsize\hfil\epsfbox{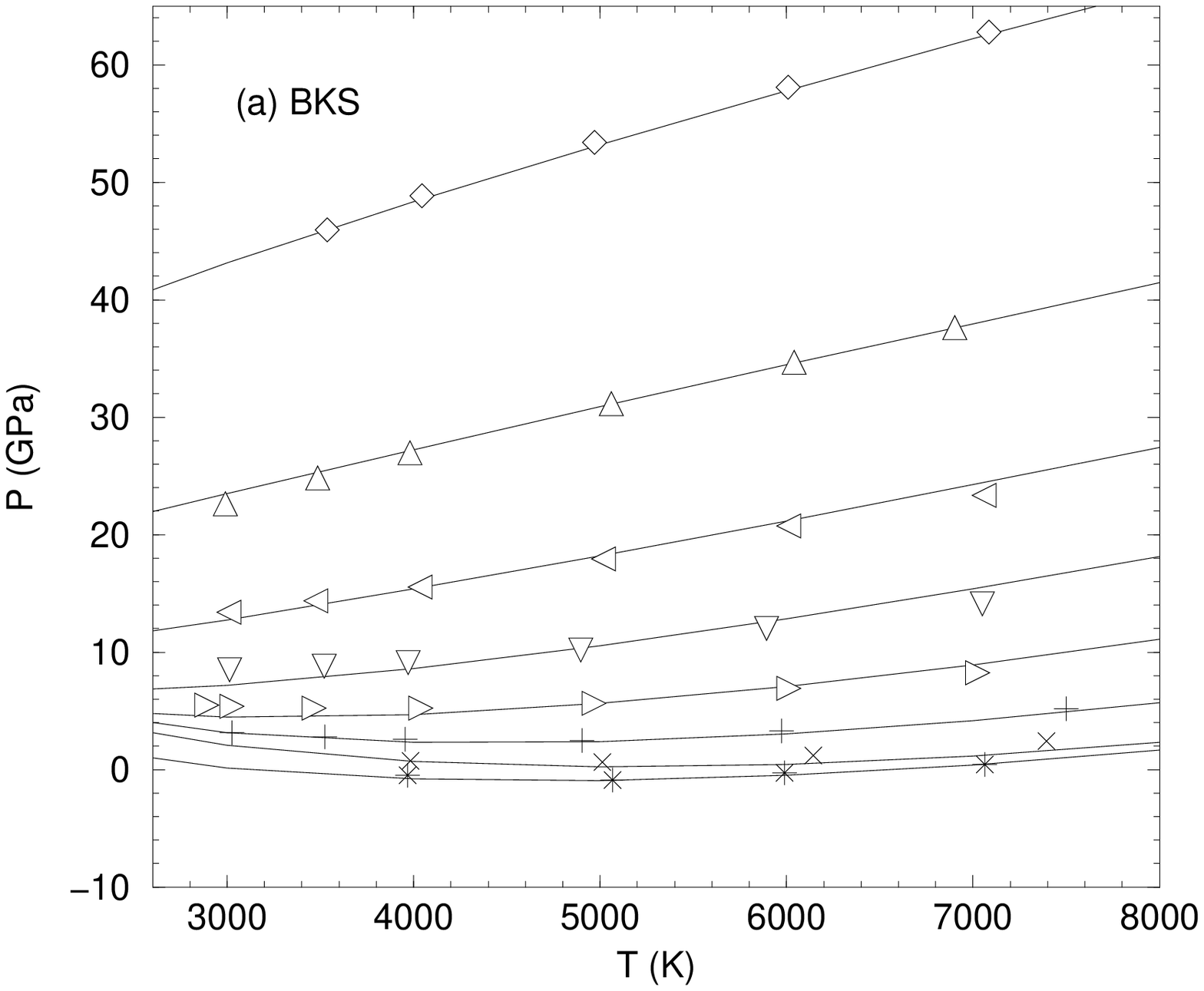}\hfil}
\hbox to\hsize{\epsfxsize=1.0\hsize\hfil\epsfbox{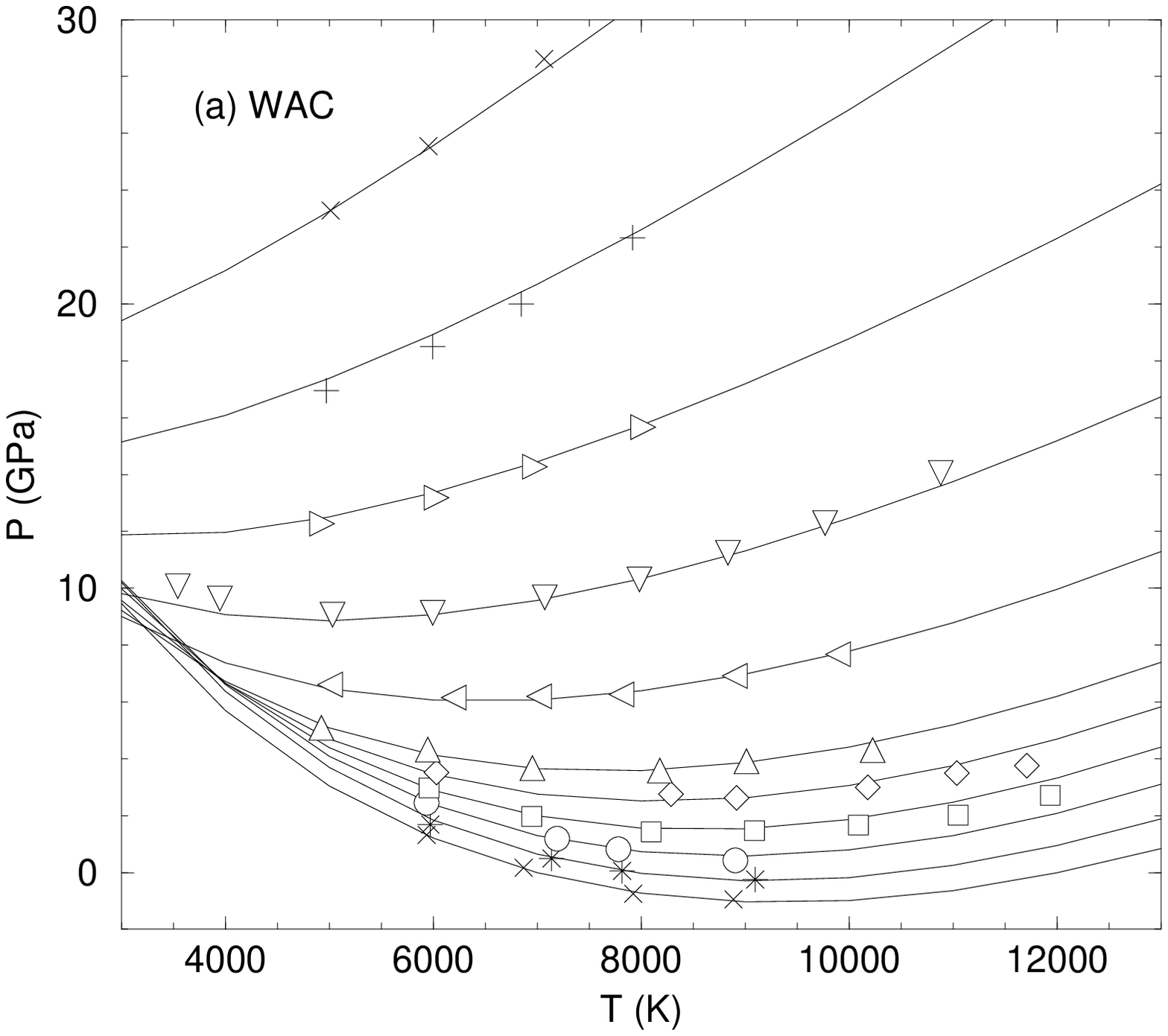}\hfil}
\caption{Isochores of $P$ versus $T$ for (a) BKS and (b) WAC silica.
Symbols are values obtained from MD simulation, while lines are
determined from the model equations of state developed in
Section~\protect{\ref{model}}.  For BKS, isochores B1 through B8 are
shown from bottom to top; for WAC, W4 through W15 are shown, from
bottom to top.}
\label{PT}
\end{figure}

To summarize, we construct a model $P(V,T)$ equation of state using as
input, polynomial fits of (i) the $V$ dependence of $a$ and $b$, and
(ii) one reference isotherm of $P$.  As a check of this equation of
state, we compare in Fig.~\ref{PT} isochores of $P$ versus $T$,
evaluated directly from simulation, and as calculated from the above
modeling procedure.

\section{Thermodynamic behavior of WAC and BKS silica}
\label{thermo}

For the description of the thermodynamic properties of tetrahedrally
coordinated liquids such as silica or water, it is useful to determine
the location and shape of curves in the space of $P$, $V$ and $T$ at
which specific thermodynamic conditions are met.  In the present
context, three such curves are important:

(i) Along the ``temperature of maximum density'' (TMD) line the
condition,
\begin{equation}
\biggr(\frac{\partial P}{\partial T}\biggl)_V=0,
\end{equation}
is satisfied~\cite{P98}.  At such a point, an isobar of $\rho$ as a
function of $T$ is a maximum, and at lower $T$, $\rho$ decreases as
$T$ decreases.  The presence of a TMD line is a hallmark of liquids in
which local tetrahedral order is prominent, and is observed
experimentally in silica, as well as in water.

(ii) The metastability limit of the liquid, or spinodal line, is
defined by~\cite{P98},
\begin{equation}
\biggr(\frac{\partial P}{\partial V}\biggl)_T=0.
\end{equation}

(iii) Along the ``$K_T^{\rm max}$ line,'' the isothermal
compressibility $K_T$ is a maximum with respect to $V$ at constant
$T$.  It is found by locating points satisfying,
\begin{equation}
\biggr(\frac{\partial K_T}{\partial V}\biggl)_T=0,
\label{ktmax}
\end{equation}
where,
\begin{equation}
K_T=-\frac{1}{V}\biggr(\frac{\partial V}{\partial P}\biggl)_T,
\end{equation}
and then checking to confirm that the extremum so identified is a
maximum~\cite{P98,singfree}.

Spinodal lines are necessarily associated with a second order critical
point that terminates a line of first order phase
transitions~\cite{P98}.  When a such a critical point occurs at
$T=T_c$, a $K_T^{\rm max}$ line will emanate from the critical point
for $T>T_c$.  However, the occurrence of a $K_T^{\rm max}$ line does
not imply the occurrence of a critical point at lower $T$; i.e. the
occurrence of a $K_T^{\rm max}$ line is a necessary, but not a
sufficient condition, for the occurrence of a critical
point~\cite{singfree}.

The locations of these lines are shown in Fig.~\ref{maps} projected
onto the $V$-$T$ plane; and in Fig.~\ref{PTmaps}, projected onto the
$T$-$P$ plane.  We find that the pattern of behavior revealed is
qualitatively the same both for WAC and BKS silica, and that this is
the same pattern found also from water simulations employing the ST2,
TIP4P~\cite{SPES97} and SPC/E models~\cite{HPSS97}.

Most significant is the occurrence both for WAC and BKS silica, of a
spinodal line in the low $T$ liquid regime that is distinct from the
liquid-gas spinodal boundary.  This spinodal is the metastability
boundary associated with a liquid-liquid phase transition.  As $T\to
0$ it coincides with the points of inflection in the $a$ versus $V$
curves of Fig.~\ref{ab}.  The critical point of this liquid-liquid
phase transition occurs at the point of maximum $T$ on the spinodal
line.  

\section{Liquid-liquid phase separation}
\label{split}

If the spinodal curves predicted by the equations of state developed
in Section~\ref{model} are correct, and if we can conduct equilibrium
simulations in the unstable regions so identified, we should observe
characteristic signs of phase separation.  When a liquid-liquid phase
separation occurs in a constant-$V$ simulation such as employed here,
both phases coexist within the simulation box, each in its own
distinct region, separated by an interface.  Each phase will have a
distinct bulk density as well as a distinct local structure.

\begin{figure}
\hbox to\hsize{\epsfxsize=1.0\hsize\hfil\epsfbox{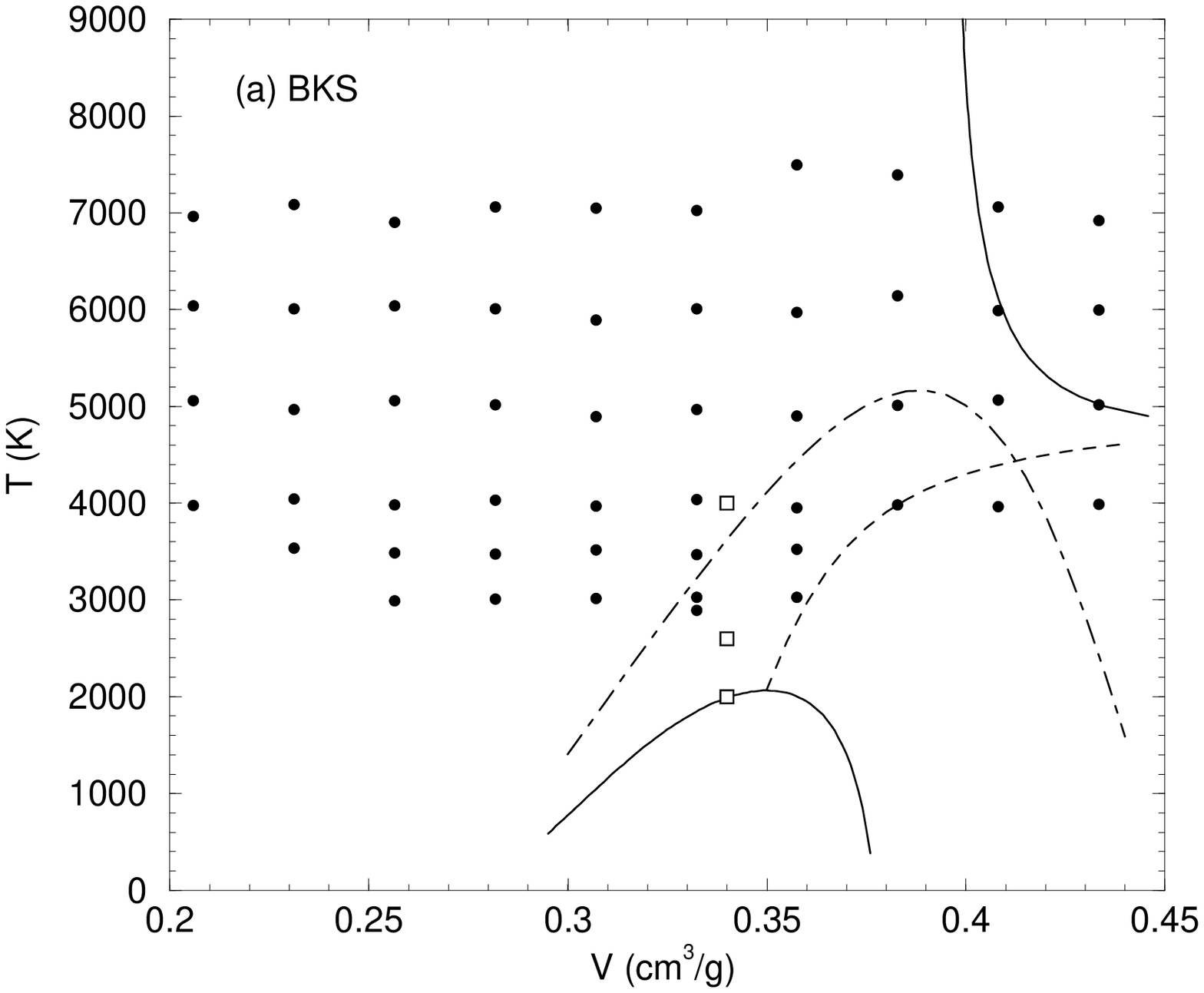}\hfil}
\hbox to\hsize{\epsfxsize=1.0\hsize\hfil\epsfbox{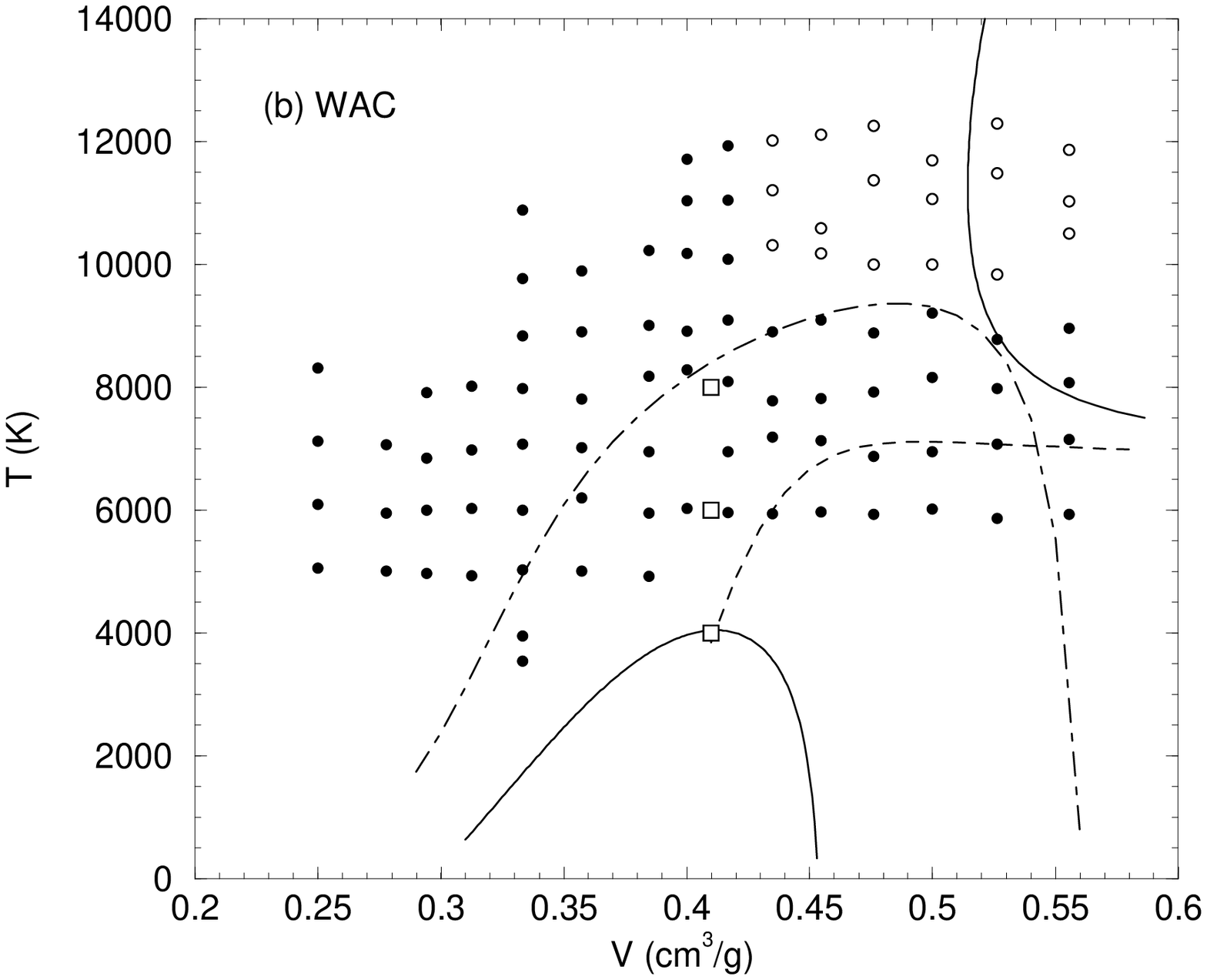}\hfil}
\caption{Simulated state points and thermodynamic features of (a) BKS
and (b) WAC silica.  The model equations of state derived in
Section~\protect{\ref{model}} are fit to MD simulation results
obtained at the $(V,T)$ points indicated by filled circles.  In (b)
open circles are WAC state points at which substantial deviations from
Eq.~\protect{\ref{tara1}} are observed, and so are excluded from the
data set used to construct the equation of state.  The BKS and WAC
equations of state give estimates for the projection into the $(V,T)$
plane of the spinodals (solid lines), TMD line (dot-dashed), and
$K_T^{max}$ line (dashed), as defined in
Section~\protect{\ref{thermo}}.  The open squares locate $(V,T)$
points at which we test for liquid-liquid phase separation, as
described in Section~\protect{\ref{split}}.}
\label{maps}
\end{figure}

We therefore carry out new MD runs for both WAC and BKS silica at
state points that approach the spinodal curve associated with the
liquid-liquid phase transition.  These state points are identified as
squares in Fig.~\ref{maps}.  To facilitate comparison of BKS and WAC
silica, we simulate a system of $N=750$ atoms for both.  The lowest
$T$ state simulated for each model ($T=2000$~K for BKS, $T=4000$~K for
WAC) is near the predicted critical point.  Because of the low $T$, we
are are unable to bring these lowest $T$ states fully into
equilibrium; however, the results obtained do serve to establish the
trend in the behavior at the lowest $T$.

To test for the occurrence of two phases with distinct local
structure, we examine the local coordination environment of the
silicon atoms.  We consider $g(r)$, the Si-Si radial distribution
function (RDF); $4\pi r^2 g(r) dr$ is the probability that a Si atom
will be found at a distance between $r$ and $r+dr$ of a reference Si
atom.  We decompose $g(r)$ according to the contributions made by
successive nearest neighbors (nn) of a given Si atom, labelled in
ascending order of distance from an Si atom.  That is, we define
sub-RDF's $g_i(r)$ according to
\begin{equation}
g(r)=\sum_{i=1}^{\infty} g_i(r),
\end{equation}
where $4\pi r^2 g_i(r) dr$ is the probability that the $i$th nearest
neighbor of a randomly selected Si atom will be found at a distance
between $r$ and $r+dr$.

\begin{figure}
\hbox to\hsize{\epsfxsize=1.0\hsize\hfil\epsfbox{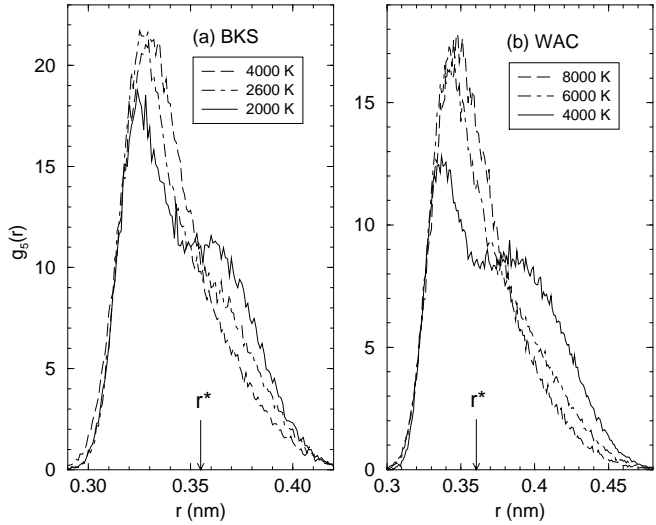}\hfil}
\caption{$g_5(r)$ for (a) BKS and (b) WAC silica evaluated at the
state points indicated by open squares in Fig.~\protect{\ref{maps}}.
The system size used is $N=750$ atoms.  The arrows indicate the values
of $r^\ast$ used to identify silicon atoms having high-density
coordination or low-density coordination.  For BKS silica,
$r^\ast=0.355$~nm; for WAC silica, $r^\ast=0.360$~nm.}
\label{gr5}
\end{figure}

Fig~\ref{gr5} shows $g_5(r)$ for BKS and WAC silica for several
different $T$ along the isochores indicated by open squares in
Fig.~\ref{maps}.  For $T$ above $T_c$, $g_5(r)$ is a unimodal function
of $r$. As $T$ decreases, the width of the $g_5(r)$ distribution
increases.  That is, rather than finding a more sharply defined 5th~nn
coordination environment as $T$ decreases, the distribution of
locations of 5th~nn's becomes broader.  For $T$ near $T_c$, $g_5(r)$
becomes bimodal.  This behavior shows that two distinct populations of
5th~nn coordination environments are emerging in the liquid as $T$
decreases.  Similar behavior is observed for $g_6(r)$, and more weakly
in $g_7(r)$ and $g_8(r)$.  This is in contrast to the behavior of
$g_1(r)$ through $g_4(r)$ which simply become sharper and narrower
distributions with decreasing $T$ (not shown).

If the emergence of these distinct coordination environments
corresponds to the onset of liquid-liquid phase separation, we should
find that Si atoms with similar coordination environments are
spatially correlated.  That is, there should be relatively compact
droplets of the two distinct phases.  

To test for this, we require systems of larger spatial extent than the
$N=750$ atom simulations used to calculate $g_5(r)$.  We therefore
initiate simulations of a system of $N=6000$ atoms for the lowest $T$
states where phase separation should be most prominent ($T=2000$~K for
BKS, $T=4000$~K for WAC).  As above, excessive equilibration times
prevent us from bringing these states fully into equilibrium, and so
we only use these simulations to establish the trend in behavior.  We
note that an incompletely equilibrated system should underestimate the
amount of phase separation, since droplets (if present) will have had
less time to form and grow.

\begin{figure}
\hbox to\hsize{\epsfxsize=1.0\hsize\hfil\epsfbox{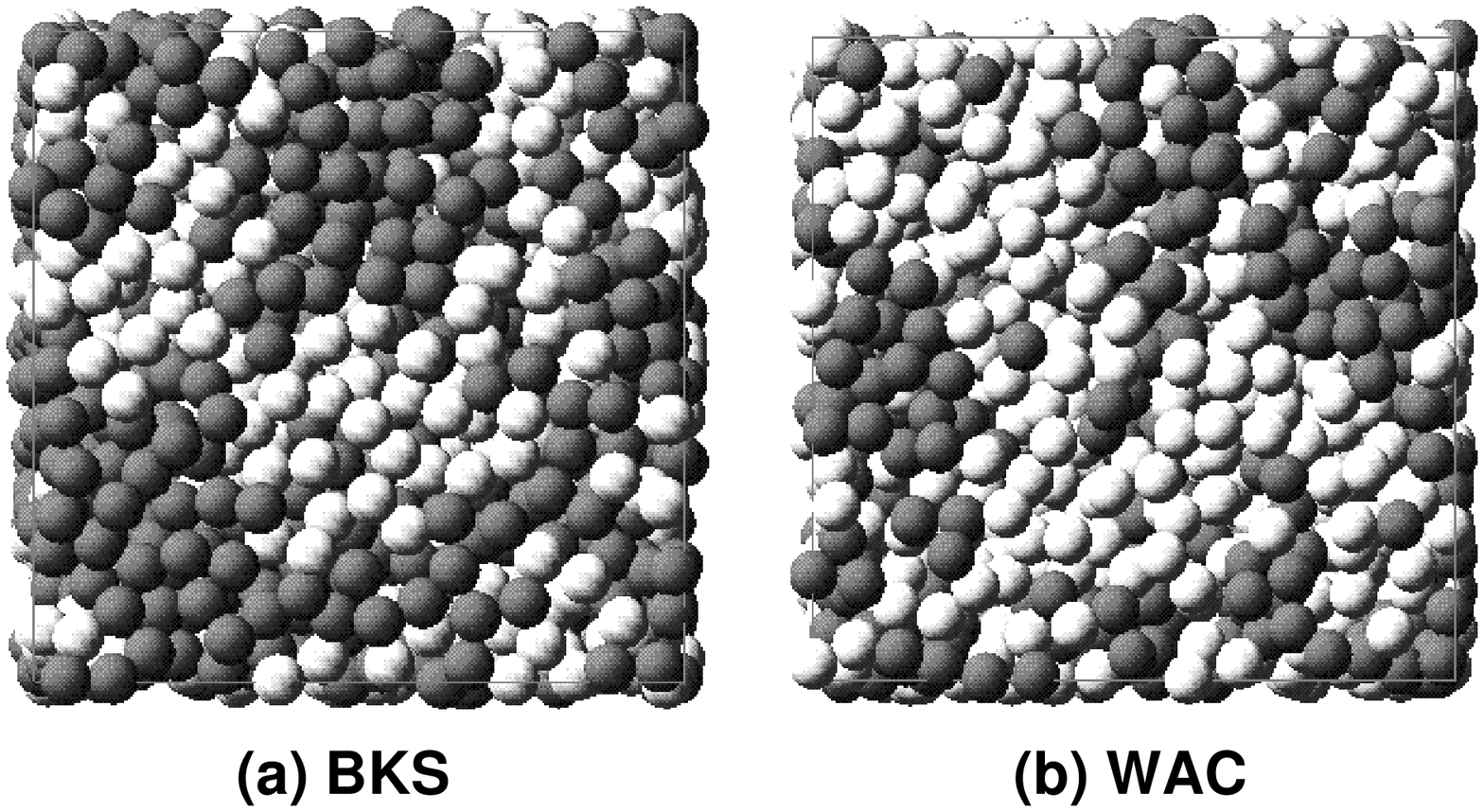}\hfil}
\caption{Snapshots of MD configurations of BKS silica ($T=2000$~K,
$\rho=0.34$~g/cm$^3$) and WAC silica ($T=4000$~K,
$\rho=0.41$~g/cm$^3$).  The configurations are viewed face-on to one
side of the simulation cell.  There are $N=6000$ atoms in each system,
but only silicon atoms are shown.  Dark spheres are silicon atoms
having high-density coordination: i.e. a 5th~nn silicon within a
distance $r<r^\ast$.  Light spheres are silicon atoms having
low-density coordination: i.e. a 5th nn silicon at a distance
$r>r^\ast$.  The clustering of atoms with similar coordination into
droplets is consistent with the onset of liquid-liquid phase
separation.}
\label{pix}
\end{figure}

We examine snapshots of these $N=6000$ atom BKS and WAC
configurations.  The minimum occurring between the two peaks of
$g_5(r)$ for the lowest $T$ in Fig.~\ref{gr5} provides a convenient
threshold $r^\ast$ for partitioning the Si atoms into two populations
according to their 5th~nn coordination.  We show in Fig.~\ref{pix}
snapshots of the positions of all the Si atoms in the BKS and WAC
systems.  Light spheres are Si atoms having their 5th~nn at a distance
greater than $r^\ast$; these atoms have a low-density coordination
environment, compared to the average. Dark spheres are Si atoms having
their 5th~nn at a distance less than $r^\ast$; these atoms have a
high-density coordination environment.  Spatially correlated droplets
of atoms with the same type of coordination environment are readily
visible, for both WAC and BKS silica.  This is consistent with the
occurrence of liquid-liquid phase separation for both of these silica
models.

\section{Discussion}

The two principal conclusions of the present work are:

(i) For two silica models, BKS and WAC, there exists a wide range of
$T$ and $V$ within which isochores of $U$ for the liquid phase conform
to Eq.~\ref{tara1}. This occurs in spite of the fact that the
prediction of Ref.~\cite{RT98} was made for simple liquids, and not
for liquids with non-trivial local structure such as occurs in silica.
This result suggests that the physical basis for Eq.~\ref{tara1} is
quite robust and provides a valuable tool for probing the low $T$
properties of a wide range of liquid systems.

(ii) The model equation of state constructed by exploiting
Eq.~\ref{tara1} predicts the occurrence of a liquid-liquid phase
transition in BKS and WAC silica.  We confirm the presence of this
transition by direct simulations near the predicted critical point.
Thus BKS and WAC silica join the rank of simulation models for
tetrahedrally coordinated liquids in which a liquid-liquid phase
transition has been directly observed.  Whether or not such a phase
transition occurs in real liquid silica (see below), it's presence in
the behavior of BKS silica is important to note, since this model is
currently in wide use for simulation studies of silica under a variety
of conditions.

Given the common behavior found for BKS and WAC silica, it is
appropriate to inquire whether we should therefore expect to find the
same pattern of thermodynamic behavior, including a liquid-liquid
phase transition, in real silica; or whether these two models share a
common flaw that makes them unrealistic in this respect.  To attempt
to address this, we seek a basis for comparing the behavior of the BKS
and WAC models with each other and with other tetrahedral liquids,
both simulated and real.  To proceed we choose as a scaling
temperature $T^{\ast}$ the highest value of $T$ reached along the TMD
line.

As shown in Table~\ref{tab2}, the ratio $T_c/T^{\ast}$ is
approximately 0.4 for both WAC and BKS silica.  Assuming that this
ratio is also valid for real silica gives $T_c=730$~K, a temperature
well below $T_g$ for silica at $1450$~K.  Hence, if the WAC and BKS
models are representative of the thermodynamic properties of silica,
then we should not expect to directly observe an equilibrium
liquid-liquid phase transition in supercooled liquid silica.

This is in contrast to the case of water.  The ST2 model gives
$T_c/T^{\ast}=0.80$, implying that $T_c=194$~K for real water.  This
is well above $T_g=136$~K for water, though as yet still outside the
easily accessible experimental range. (See however, Ref.~\cite{MS98}.)

The present analysis is consistent with recent
interpretations~\cite{SPES97, PHA97, L00} of the phenomenology of
polyamorphism as observed in amorphous solid silica and water.  In
this interpretation, the manifestation of polyamorphism for a system
where the liquid phase exhibits the tendency toward a liquid-liquid
phase transition at low $T$ will depend on the relationship of $T_g$
to $T_c$.  When $T_c>T_g$, as may be the case for water, polyamorphic
behavior in the amorphous solid will be prominent, with a large and
sharp density increase observed when the glass in compressed.  When
$T_c<T_g$, behavior characteristic of polyamorphism will be weaker,
though not necessarily absent, as is the case for silica.

However, the precise nature of the behavior to be expected when
$T_c<T_g$ remains an open area of research.  For $T<T_g$, a standard
application of equilibrium thermodynamics is not appropriate and so
predictions based on the model equations of state presented here do
not apply.  Nonetheless, new thermodynamic approaches, based on a
separation of the configurational degrees of freedom (frozen in at
$T_g$) and vibrational degrees of freedom (which are always in thermal
equilibrium) may be employed to derive a free energy expression that
can be used used to locate the (cooling-rate dependent) location of
the critical point~\cite{teo,fd}.  In this approach, a negative
curvature of the $V$ dependence of $U$ along the Kauzmann line would
be sufficient to guarantee the presence of an instability in the
glassy phase.

\begin{figure}
\hbox to\hsize{\epsfxsize=1.0\hsize\hfil\epsfbox{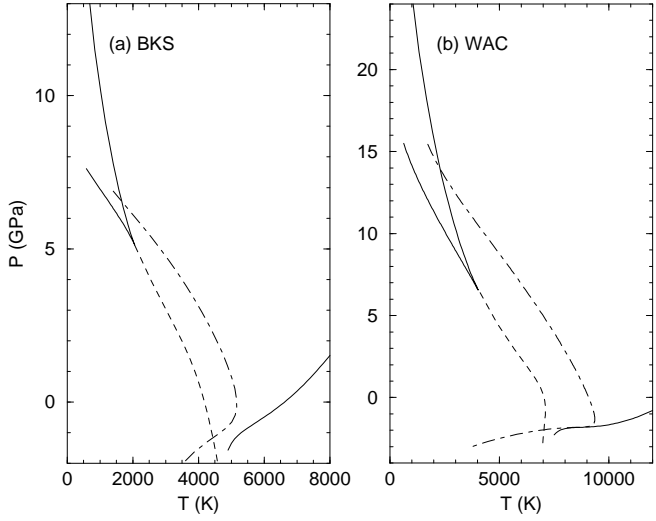}\hfil}
\caption{Estimates for the projection into the $(T,P)$ plane of the
spinodals (solid lines), TMD line (dot-dashed), and $K_T^{max}$ line
(dashed), evaluated from our model equations of state for (a) BKS and
(b) WAC silica.}
\label{PTmaps}
\end{figure}

We also note that the value of $T$ for the $K_T^{max}$ line at
atmospheric $P$ is about 80\% of $T^{\ast}$ for both WAC and BKS
silica (see Fig.~\ref{PTmaps}).  Applying this ratio to real silica
suggests that a maximum of $K_T$ (or, equivalently, a minimum of the
bulk modulus) could occur when liquid silica is compressed
isothermally at $T$ just above, but near $T_g$.  This raises the
possibility to directly observe a $K_T^{max}$ line in experiments of
supercooled liquid silica under pressure.  Confirming or refuting this
prediction would be an important step in establishing the
applicability of the pattern of thermodynamic behavior presented here,
to real silica and other tetrahedral liquids.

\begin{acknowledgments}
We are grateful to G. Parisi for useful discussions.  ISV and PHP wish
to thank NSERC (Canada) for financial support, and the Dipartimento di
Fisica and Istituto Nazionale per la Fisica della Materia, Universita'
di Roma La Sapienza, for their financial support and hospitality
during visits when portions of this work were carried out.  FS
acknowledges financial support from INFM-PRA-HOP and PRIN98.
Substantial computing resources were provided by the Compaq-Western
Centre for Computational Research.
\end{acknowledgments}


\bigskip

\begin{table}
\begin{tabular}{|l|c|c||l|c|c|}
\hline 
\multicolumn{3}{|c||}{BKS isochores} &
\multicolumn{3}{|c|}{WAC isochores}
\\ \hline

  label & $V$ (cm$^3$/g) & $\rho$ (g/cm$^3$) 
& label & $V$ (cm$^3$/g) & $\rho$ (g/cm$^3$) \\ \hline

B1 &  0.4334 & 2.3071 & W1  & 0.5555  &  1.8\\
B2 &  0.4081 & 2.4501 & W2  & 0.5263  &  1.9\\
B3 &  0.3828 & 2.6119 & W3  & 0.5000  &  2.0\\
B4 &  0.3575 & 2.7966 & W4  & 0.4761  &  2.1\\
B5 &  0.3322 & 3.0093 & W5  & 0.4545  &  2.2\\
B6 &  0.3070 & 3.2572 & W6  & 0.4347  &  2.3\\
B7 &  0.2817 & 3.5495 & W7  & 0.4166  &  2.4\\
B8 &  0.2564 & 3.8994 & W8  & 0.4000  &  2.5\\
B9 &  0.2311 & 4.3260 & W9  & 0.3846  &  2.6\\
B10&  0.2058 & 4.8572 & W10 & 0.3571  &  2.8 \\ 
   &        &         & W11 & 0.3333  &  3.0 \\
   &        &         & W12 & 0.3125  &  3.2 \\
   &        &         & W13 & 0.2941  &  3.4 \\
   &        &         & W14 & 0.2777  &  3.6 \\
   &        &         & W15 & 0.2500  &  4.0 \\ \hline

\end{tabular}
\caption{Volume $V$ and density $\rho$ of each isochore simulated for
BKS and WAC silica.  The labels are used to identify isochores shown
in the figures.}
\label{vlist}
\end{table}

\bigskip

\begin{table}
\begin{tabular}{|c|c|c|c|c|} 
\hline    
model   & n & $\alpha_n$ 
        & $\beta_n$ 
        & $\gamma_n$ \\ \hline

WAC           & 0 & -9.13919 & -0.00627597 &  122.772 \\ 
$T_0=7000$~K  & 1 & -38.8581 &  0.108913   & -328.537 \\    
              & 2 &  163.650 & -0.494217   &  311.922 \\    
              & 3 & -306.599 &  0.983621   & -328.537 \\    
              & 4 &  205.422 & -0.683688   &  29.1778 \\   
              & 5 &  -       &  -          &  2.31542 \\ \hline    

BKS           & 0 & -1.01756 &  0.00144202 &  479.401   \\
$T_0=5000$~K  & 1 & -61.1632 &  0.0488104  & -708.648   \\   
              & 2 &  293.988 & -0.381123   &  409.488   \\   
              & 3 & -621.091 &  1.069764   & -117.045   \\    
              & 4 &  478.866 & -0.969089   &  16.6264   \\   
              & 5 &  -       &  -          & -0.904994  \\    
\hline    
\end{tabular}
\caption{Coefficients of polynomial fits for $V$ dependence of $a$ and
$b$; and the $\rho$ dependence of $P$ for $T=T_0$.  The units for each
coefficient are appropriate to give $a$ in MJ/mol, $b$ in
MJ/(mol\,T$^{3/5}$), and $P$ in GPa, when $V$ is given in cm$^3$/g and
$\rho$ is measured in g/cm$^3$.
\label{coeffs}}
\end{table}

\bigskip

\begin{table}
\begin{tabular}{|l|c|c|c|c|c|c|}
\hline & $T^{\ast}$ & $T_c$ & $T_g$ & ${{T_g}/{T^{\ast}}}$ &
${{T_c}/{T^{\ast}}}$ & ${{T_c}/{T_g}}$ \\ \hline WAC & 9000 & 4000
&$<$4000 & $<$0.44 & 0.44 & $>$1 \\ BKS & 5000 & 2000 & 1380~\cite{HK}
& 0.28 & 0.40 & 1.45 \\ ST2 & 330~\cite{PSES92} & 235~\cite{HZPSS97} &
$<$235 & $<$0.71 & 0.71 & $>$1 \\ \hline SiO$_2$ &
1823~\cite{silicaTMD} & $\sim 730$ & 1450~\cite{silicaTg} & 0.80 &
$\sim 0.40$ & 0.50 \\ H$_2$O & 277~\cite{waterTMD} & $\sim 194$ &
136~\cite{waterTg} & 0.49 & $\sim 0.70$ & 1.4 \\ \hline
\end{tabular}
\caption{Comparison of characteristic temperatures of simulated and
real tetrahedral liquids. $T^{\ast}$ for real SiO$_2$ and H$_2$O are
estimated as the ambient $P$ values of the TMD; the actual value of
$T^{\ast}$ for these systems is likely to be slightly higher.  Upper
bounds on $T_g$ for WAC silica and ST2 water are taken simply as the
lowest $T$ at which equilibrium simulations of the liquid have been
conducted.  Values preceded by ``$\sim$'' are found by assuming that
the values of ${{T_c}/{T^{\ast}}}$ found from simulations apply to the
real substances.  References are provided in brackets for those values
not determined here.
\label{tab2}}
\end{table}


\begin{thebibliography}{99}

\bibitem{T84} M.O. Thompson, G.J. Galvin, J.W. Mayer, P.S. Peercy,
J.M. Poate, D.C Jacobson, A.G. Cullis and N.G. Chew,
Phys. Rev. Lett. {\bf 52}, 2360 (1984).

\bibitem{DSTPJ85} E.P. Donovan, F. Saepen, D. Turnbull, J.M. Poate,
and D.C. Jacobson, J. Appl. Phys. {\bf 57}, 1795 (1985).

\bibitem{BVPU91} V.V. Brazhkin, R.N. Voloshin, S.V. Popova and
A.G. Umnov, Phys. Lett. A {\bf 154}, 413 (1991).

\bibitem{BPVU92} V.V. Brazhkin, S.V. Popova, R.N. Voloshin and
A.G. Umnov, High Pressure Research {\bf 6}, 363 (1992).

\bibitem{AM94} S. Aasland and P.F. McMillan, Nature {\bf 369}, 633
(1994).

\bibitem{T97} M. Togaya, Phys. Rev. Lett. {\bf 79}, 2474 (1997).

\bibitem{MS98} O. Mishima and H.E. Stanley, Nature {\bf 396}, 329
(1998).

\bibitem{M00} O. Mishima, Phys. Rev. Lett. {\bf 85}, 334
(2000).

\bibitem{K00} Y. Katayama, T. Mizutani, W. Utsumi, O. Shimomura,
M. Yamakata and K.-I. Funakoshi, Nature {\bf 403}, 170 (2000).

\bibitem{LL89} W.D. Ludke and U. Landman, {Phys. Rev. B} {\bf 37},
4656 (1988)

\bibitem{ABG96} C.A. Angell, S. Borick and M. Grabow,
J. Non-Cryst. Solids {\bf 205-207}, 463 (1996).

\bibitem{PSES92} P.H.~Poole, F.~Sciortino, U.~Essmann and
H.~E.~Stanley, Nature {\bf 360}, 324 (1992).

\bibitem{HZPSS97} S. Harrington, R. Zhang, P. H. Poole, F. Sciortino
and H. E. Stanley, Phys. Rev. Lett. {\bf 78}, 2409 (1997).

\bibitem{GR99} J.N. Glosli and F.H. Ree, Phys. Rev. Lett. {\bf 82},
4659 (1999).

\bibitem{HS70} P.C. Hemmer and G. Stell,  Phys. Rev. Lett. {\bf 24},
1284 (1970).

\bibitem{MPS85} A.C. Mitus, A.Z. Patashinskii and B.I. Shumilo,
Phys. Lett. {\bf 113A}, 41 (1985).

\bibitem{PB92} E.G.  Ponyatovsky and I.O. Barkalov, Materials Science
Reports {\bf 8}, 147 (1992).

\bibitem{SSS93}  S. Sastry, F. Sciortino and H.E. Stanley,
J. Chem. Phys. {\bf 98}, 9863, (1993).

\bibitem{PSGSA94} P.H. Poole, F. Sciortino, T. Grande, H.E. Stanley
and C.A. Angell, Phys. Rev. Lett. {\bf 73}, 1632 (1994).

\bibitem{RPD96} C. J. Roberts, A. Z. Panagiotopoulos and Pablo
G. Debenedetti, Phys. Rev. Lett. {\bf 77}, 4386 (1996).

\bibitem{TB98} C.F. Tejero and Marc Baus, Phys. Rev. E {\bf 57}, 4821
(1998).

\bibitem{J99} E.A. Jagla, J. Chem. Phys. {\bf 111}, 8980 (1999).

\bibitem{FMSBS00} G. Franzese, et al., preprint, cond-mat/0005184.

\bibitem{J00} J.A. Jagla, preprint, cond-mat/0006381.

\bibitem{W92} G.H. Wolf, S. Wang, C.A. Herbst, D.J. Durben,
W.J. Oliver, Z.C. Kang, and C. Halvorsen, in {\it High--Pressure
Research: Application to Earth and Planetary Sciences}, edited by
Y.S. Manghnani and M.H. Manghnani (American Geophysical Union,
Washington, 1992) p.~503.

\bibitem{PGAM97} P.H. Poole, T. Grande, C.A. Angell and P.F. McMillan
Science {\bf 275}, 322 (1997).  

\bibitem{MCW85} O. Mishima, L.D. Calvert and E. Whalley, Nature {\bf
314}, 76 (1985).

\bibitem{MTA91} O. Mishima, K. Takemura and K. Aoki, Science {\bf 254},
406 (1991).

\bibitem{M94} O. Mishima, J. Chem. Phys. {\bf 100}, 5910 (1994).

\bibitem{SR74} F.H. Stillinger and A. Rahman, J. Chem. Phys. {\bf 60},
1545 (1974).

\bibitem{PESS93} P.H.~Poole, U.~Essmann, F.~Sciortino and
H.~E.~Stanley, Phys. Rev. E {\bf 48}, 4605 (1993).

\bibitem{SPES97} F. Sciortino, P.H. Poole, U. Essmann, and
H.E. Stanley, Phys. Rev. E {\bf 55}, 727 (1997).

\bibitem{JCMIK83} W.L. Jorgensen, J. Chandrasekhar, J. Madura, R.W.
Impey and M. Klein, J. Chem. Phys. {\bf 79}, 926 (1983).   

\bibitem{B84} H.J.C.Berendsen, J.P.M.Postma, W.F.Van Gunsteren,
A.Dinola, J.R.Haak, J. Chem. Phys. {\bf81}, 3684 (1984)

\bibitem{HPSS97} S. Harrington, P. H. Poole, F. Sciortino and
H. E. Stanley, J. Chem. Phys. {\bf 107}, 7443 (1997)

\bibitem{P98} P.G. Debenedetti, {\it Metastable Liquids: Concepts and
Principles}, Princeton University Press, Princeton (1998).

\bibitem{SKS82} H. Sugiura, K.--I. Kondo and A. Sawaoka in {\it
High--Pressure Research in Geophysics} (eds S. Akimoto and
M.H. Manghnani) 551 (Reidel, Dordrecht, 1982) 

\bibitem{G84} M. Grimsditch, {Phys. Rev. Lett.} {\bf 52}, 2379 (1984).

\bibitem{HMBM86} R. J. Hemley, H. K. Mao, P. M. Bell, and B. O. Mysen
Phys. Rev. Lett. {\bf 57}, 747 (1986).

\bibitem{SSGCS93} M.S. Somayazulu, et al. {J. Phys.: Condens. Matter}
{\bf 5}, 6345 (1993).

\bibitem{TKL92} J.S. Tse, D.D. Klug, and Y. LePage, {Phys. Rev. B}
{\bf 46}, 5933 (1992).

\bibitem{JKV93} W. Jin, R.K. Kalia, P. Vashishta and J.P. Rino,
{Phys. Rev. Lett.} {\bf 71}, 3146 (1993).

\bibitem{PHA97} P.H. Poole, M. Hemmati and C.A. Angell
Phys. Rev. Lett. {\bf 79}, 2281 (1997).
        
\bibitem{L00} D.L. Jacks, Phys. Rev. Lett. {\bf 84}, 4629 (2000).

\bibitem{WAC76} L.V. Woodcock, C.A. Angell and P.A. Cheeseman,
{J. Chem. Phys.} {\bf 65}, 1565 (1976).

\bibitem{RT98} Y. Rosenfeld and P. Tarazona, Mol. Phys. {\bf 95}, 141
(1998).

\bibitem{prlentro}  F. Sciortino, W. Kob and P. Tartaglia,
Phys. Rev. Lett.  {\bf 83}, 3214 (1999).

\bibitem{barbara} B. Coluzzi, Ph.D Thesis, University of Roma La
Sapienza (1999); B. Coluzzi, P. Verrocchio and G. Parisi,
Phys. Rev. Lett.  {\bf 84}, 306 (2000)

\bibitem{srinew} S. Sastry, Phys. Rev. Lett. {\bf 85}, 590 (2000).

\bibitem{BKS90} B.W.H. van Beest, G.J. Kramer, and R.A. van
Santen, Phys. Rev. Lett. {\bf 64}, 1955 (1990).

\bibitem{AT89} M.P. Allen and D.J. Tildesley, {\it Computer Simulation
of Liquids} (Oxford University Press, Oxford, 1989).

\bibitem{mod} To correct for the unphysical behavior of the BKS
model at short range, we use a potential of the form,
$\Phi_{ij}=\Phi^{BKS}_{ij} + 4 \epsilon_{ij} [ (\sigma_{ij}/r)^{30} -
(\sigma_{ij}/r)^6 ]$, where $\Phi^{BKS}_{ij}$ is the original BKS
form, and $r$ is the separation between between atoms $i$ and $j$,
which may be either Si or O atoms. $\epsilon_{SiSi}=\sigma_{SiSi}=0$,
while $\epsilon_{SiO}=4.9634598\times 10^{-22}$~J,
$\epsilon_{SiO}=0.1313635$~nm, $\epsilon_{OO}=1.6839685\times
10^{-22}$~J, and $\sigma_{OO}=0.1779239$~nm.  $\sigma_{ab}$ and
$\epsilon_{ab}$ are chosen so that the original BKS potential is
modified as little as possible at larger $r$, subject to the
constraint that the new potential has no inflection at small $r$.

\bibitem{Callen} H.B. Callen, {\it Thermodynamics and an Introduction
to Thermostatistics,} 2nd edition (John Wiley and Sons, New York,
1985).

\bibitem{maple} The curve fitting, differentiations, and integrations
required to build the functional model of the thermodynamic properties
are carried out using the symbolic computation system {\it Maple},
version V.5, Waterloo Maple, Inc., Waterloo, Canada.

\bibitem{singfree} S. Sastry, P.G. Debenedetti, F. Sciortino, and
H.E. Stanley, { Phys. Rev. E} {\bf 53}, 6144 (1996).  Note that the
definition used here for the $K_T^{\rm max}$ line differs from that of
the ``temperature of extremal compressibility'' line studied by these
authors.  We focus on the $K_T^{\rm max}$ line because it's location
would be simpler to identify in isothermal compression experiments.

\bibitem{teo} Th. M. Nieuwenhuizen, { Phys. Rev. Lett.} {\bf 80}, 5580
(1998); {\it ibid.} {\bf 79}, 1317 (1997).

\bibitem{fd} F. Sciortino and P. Tartaglia, cond-mat/0007208 (2000).

\bibitem{HK} J. Horbach and W. Kob, Phys. Rev. B {\bf 60}, 3169
(1999).

\bibitem{silicaTMD} C.A. Angell and H. Kanno, { Science} {\bf 193},
1121 (1976); C.A. Angell, P.A. Cheeseman and S. Tamaddam, { Science}
{\bf 218}, 885 (1982).

\bibitem{silicaTg} R. Bruckner, J. Non-Cryst. Solids {\bf 5}, 123
(1970).

\bibitem{waterTMD} R.A. Fine and F.J. Millero, J. Chem. Phys. {\bf
59}, 5529 (1973).

\bibitem{waterTg} G.P. Johari, A. Hallbrucher and E. Mayer, Nature
{\bf 330}, 552 (1987).

\end{thebibliography}
\end{document}